\documentclass[runningheads]{llncs}

\authorrunning{T. Nipkow and S. Roßkopf}

\usepackage{isabelle,isabellesym}
\isabellestyle{it}

\renewcommand{\isakeyword}[1]
{{\normalfont\sffamily\bfseries\def\isachardot{.}\def\isacharunderscore{\isacharunderscorekeyword}%
\def\isacharbraceleft{\{}\def\isacharbraceright{\}}#1}}
\newcommand{\noquotes}[1]{{\renewcommand{\isachardoublequote}{}\renewcommand{\isachardoublequoteopen}{}\renewcommand{\isachardoublequoteclose}{}#1}}

\usepackage{amssymb}
\usepackage{mathpartir}
\usepackage{url}
\usepackage[hidelinks]{hyperref}

\newcommand{\M}{$\mathcal{M}$}

\newcommand{\withquotes}[1]{\renewcommand{\isachardoublequote}{"}{#1}}

\begin{document}

\title{Isabelle's Metalogic:\\ Formalization and Proof Checker\thanks{Supported by
    Wirtschaftsministerium Bayern under DIK-2002-0027//DIK0185/03 and DFG
    GRK 2428 ConVeY}}

\author{Tobias Nipkow\orcidID{0000-0003-0730-515X} \and Simon Roßkopf\orcidID{0000-0002-7955-8749}}

\institute{Technical University of Munich, Germany}

\maketitle

\begin{isabellebody}%
\setisabellecontext{LaTeXsugar}%
\isadelimtheory
\isanewline
\endisadelimtheory
\isatagtheory
\endisatagtheory
{\isafoldtheory}%
\isadelimtheory
\endisadelimtheory
\isadelimML
\endisadelimML
\isatagML
\endisatagML
{\isafoldML}%
\isadelimML
\endisadelimML
\isadelimtheory
\endisadelimtheory
\isatagtheory
\endisatagtheory
{\isafoldtheory}%
\isadelimtheory
\endisadelimtheory
\end{isabellebody}%

\begin{isabellebody}%
\setisabellecontext{Paper}%
\isadelimtheory
\endisadelimtheory
\isatagtheory
\endisatagtheory
{\isafoldtheory}%
\isadelimtheory
\endisadelimtheory
\isadelimproof
\endisadelimproof
\isatagproof
\endisatagproof
{\isafoldproof}%
\isadelimproof
\endisadelimproof
\isadelimproof
\endisadelimproof
\isatagproof
\endisatagproof
{\isafoldproof}%
\isadelimproof
\endisadelimproof
\isadelimproof
\endisadelimproof
\isatagproof
\endisatagproof
{\isafoldproof}%
\isadelimproof
\endisadelimproof
\isadelimproof
\endisadelimproof
\isatagproof
\endisatagproof
{\isafoldproof}%
\isadelimproof
\endisadelimproof
\isadelimproof
\endisadelimproof
\isatagproof
\endisatagproof
{\isafoldproof}%
\isadelimproof
\endisadelimproof
\isadelimproof
\endisadelimproof
\isatagproof
\endisatagproof
{\isafoldproof}%
\isadelimproof
\endisadelimproof
\isadelimproof
\endisadelimproof
\isatagproof
\endisatagproof
{\isafoldproof}%
\isadelimproof
\endisadelimproof
\isadelimproof
\endisadelimproof
\isatagproof
\endisatagproof
{\isafoldproof}%
\isadelimproof
\endisadelimproof
\isadelimproof
\endisadelimproof
\isatagproof
\endisatagproof
{\isafoldproof}%
\isadelimproof
\endisadelimproof
\isadelimproof
\endisadelimproof
\isatagproof
\endisatagproof
{\isafoldproof}%
\isadelimproof
\endisadelimproof
\isadelimproof
\endisadelimproof
\isatagproof
\endisatagproof
{\isafoldproof}%
\isadelimproof
\endisadelimproof
\isadelimproof
\endisadelimproof
\isatagproof
\endisatagproof
{\isafoldproof}%
\isadelimproof
\endisadelimproof
\isadelimproof
\endisadelimproof
\isatagproof
\endisatagproof
{\isafoldproof}%
\isadelimproof
\endisadelimproof
\isadelimproof
\endisadelimproof
\isatagproof
\endisatagproof
{\isafoldproof}%
\isadelimproof
\endisadelimproof
\isadelimproof
\endisadelimproof
\isatagproof
\endisatagproof
{\isafoldproof}%
\isadelimproof
\endisadelimproof
\begin{isamarkuptext}%
\begin{abstract}
Isabelle is a generic theorem prover with a fragment of higher-order
logic as a metalogic for defining object logics. Isabelle also
provides proof terms. We formalize this metalogic and the language of
proof terms in Isabelle/HOL, define an executable (but inefficient)
proof term checker and prove its correctness w.r.t.\ the metalogic.
We integrate the proof checker with Isabelle and
run it on a range of logics and theories to check the correctness of all the
proofs in those theories.
\end{abstract}

\section{Introduction}

One of the selling points of proof assistants is their
trustworthiness. Yet in practice soundness problems do come up in
most proof assistants. Harrison \cite{Harrison-holhol} distinguishes errors
in the logic and errors in the implementation (and cites examples).
Our work contributes to the solution of both problems for the proof
assistant Isabelle \cite{Isabelle-LNCS}. Isabelle is a
generic theorem prover: it implements \M, a fragment of intuitionistic
higher-order logic, as a metalogic for defining object logics.
Its most developed object logic is HOL and the resulting proof
assistant is called Isabelle/HOL \cite{LNCS2283,Concrete}. The latter is the
basis for our formalizations.

Our first contribution is the first complete formalization of Isabelle's
metalogic. Thus our work applies to all Isabelle
object logics, e.g. not just HOL but also ZF. Of course Paulson
\cite{Paulson-JAR-89} describes \M\ precisely, but only on paper. More
importantly, his description does not cover polymorphism and type classes, which
were introduced later \cite{cade/NipkowP92}.  The published account of
Isabelle's proof terms \cite{BerghoferN-TPHOLs00} is also silent about
type classes. Yet type classes are a significant complication (as, for example,
Kun\v{c}ar and Popescu \cite{jar/KuncarP19a} found out).

Our second contribution is a verified (against \M) and executable
checker for Isabelle's proof terms. We have integrated the proof
checker with Isabelle. Thus we can guarantee that every theorem 
whose proof our proof checker accepts is provable in our definition 
of \M.  So far we are able to check the correctness of moderatly sized
theories across the full range of logics implemented in Isabelle.

Although Isabelle follows the LCF-architecture (theorems that can only
be manufactured by inference rules) it is based on an infrastructure
optimized for performance. In particular, this includes multithreading, which is
used in the kernel and has once lead to a soundness issue\footnote{\scriptsize\url{https://mailmanbroy.in.tum.de/pipermail/isabelle-dev/2016-December/007251.html}}
.
Therefore we opt for the ``certificate
checking'' approach (via proof terms) instead of verifying the implementation.

This is the first work that deals directly with what is
implemented in Isabelle as opposed to a study of the metalogic that
Isabelle is meant to implement.  Instead of reading the implementation
you can now read and build on the more abstract formalization in this paper. The
correspondence of the two can be established for each proof by
running the proof checker.

Our formalization reflects the ML implementation of Isabelle's terms
and types and some other data structures.
Thus a few implementation choices are visible, e.g.\
De Bruijn indices. This is necessary because we want to
integrate our proof checker as directly as possible with Isabelle,
with as little unverified glue code as possible, for example no
translation between De Bruijn indices and named variables.
We refer to this as our \emph{intentional implementation bias}.
In principle, however, one could extend our formalization with
different representations (e.g.\ named terms) and prove suitable
isomorphisms.
Our work is purely proof theoretic; semantics is out of scope.

\subsection{Related Work}

Harrison \cite{Harrison-holhol} was the first to verify some of HOL's metatheory and an
implementation of a HOL kernel in HOL itself. Kumar \emph{et al.}
\cite{jar/KumarAMO16} formalized HOL including definition principles,
proved its soundness and synthesized a verified kernel of a HOL prover
down to the machine language level. Abrahamsson \cite{jlap/Abrahamsson20} verified
a proof checker for the OpenTheory \cite{Hurd-PLMMS09} proof
exchange format for HOL.

Wenzel \cite{tphol/Wenzel97} showed how to interpret type classes as predicates on types.
We follow his approach of reflecting type classes in the logic but cannot remove them completely
because of our intentional implementation bias (see above).
Kun\v{c}ar and Popescu
\cite{itp/Kuncar015,esop/Kuncar017,pacmpl/Kuncar018,jar/KuncarP19a}
focus on the subtleties of definition principles for HOL with
overloading and prove that under certain conditions, type and
constant definitions preserve consistency. {\AA}man Pohjola \emph{et al.}
\cite{lpar/PohjolaG20} formalize \cite{itp/Kuncar015,jar/KuncarP19a}.

Adams~\cite{itp/Adams16} presents HOL Zero, a basic theorem prover for
HOL that addresses the problem of how to ensure that parser and
pretty-printer do not misrepresent formulas.

Let us now move away from Isabelle and HOL. Sozeau \emph{et al.}
\cite{pacmpl/SozeauBFTW20} present the first implementation of a type
checker for the kernel of Coq that is proved correct in Coq with
respect to a formal specification. Carneiro \cite{mkm/Carneiro20} has
implemented a highly performant proof checker for a multi-sorted first
order logic and is in the process of verifying it in its own logic.

We formalize a logic with bound variables, and there is a large body of related work
that deals with this issue (e.g.\ \cite{Urban-JAR08,JARsiASN,jar/GheriP20})
and a range of logics and systems with special
support for handling bound variables (e.g.\ \cite{Pfenning-LICS-89,cade/PfenningS99,flops/Pientka10}).
We found that De Bruijn indices worked reasonably well for us.

\section{Preliminaries}

Isabelle types are built from type variables, e.g. \isa{{\isacharprime}a}, and (postfix) type constructors,
e.g. \isa{{\isacharprime}a\ list}; the function type arrow is \isa{{\isasymRightarrow}}.
Isabelle also has a type class system explained later.
The notation $t :: \tau$ means that term $t$ has type $\tau$.
Isabelle/HOL provides types
\isa{{\isacharprime}a\ set} and \isa{{\isacharprime}a\ list} of sets and lists of elements
of type \isa{{\isacharprime}a}. They come with the following vocabulary: function \isa{\textsf{set}}
(conversion from lists to sets), \isa{{\isacharparenleft}{\isacharhash}{\isacharparenright}} (list
constructor), \isa{{\isacharparenleft}{\isacharat}{\isacharparenright}} (append), \isa{{\isacharbar}xs{\isacharbar}} (length of list \isa{xs}),
\isa{xs\ {\isacharbang}\ i} (the \isa{i}th element of \isa{xs} starting at 0),
\isa{\textsf{list-all2}} \isa{p\ {\isacharbrackleft}x\isactrlsub {\isadigit{1}}{\isacharcomma}\ {\isasymdots}{\isacharcomma}\ x\isactrlsub m{\isacharbrackright}\ {\isacharbrackleft}y\isactrlsub {\isadigit{1}}{\isacharcomma}\ {\isasymdots}{\isacharcomma}\ y\isactrlsub n{\isacharbrackright}\ {\isacharequal}\ {\isacharparenleft}m\ {\isacharequal}\ n\ {\isasymand}\ p\ x\isactrlsub {\isadigit{1}}\ y\isactrlsub {\isadigit{1}}\ {\isasymand}\ {\isasymdots}\ {\isasymand}\ p\ x\isactrlsub n\ y\isactrlsub n{\isacharparenright}}
and other self-explanatory notation.

The \isa{\textsf{Field}} of a relation \isa{r} is the set of all
\isa{x} such that {\renewcommand{\isacharunderscore}{\_}\isa{{\isacharparenleft}x{\isacharcomma}{\isacharunderscore}{\isacharparenright}} or \isa{{\isacharparenleft}{\isacharunderscore}{\isacharcomma}x{\isacharparenright}}} is in \isa{r}.

There is also the predefined data type %
\begin{isabelle}%
\isacommand{datatype}\ {\isacharprime}a\ option\ {\isacharequal}\ \textsf{None}\ {\isacharbar}\ \textsf{Some}\ {\isacharprime}a%
\end{isabelle}
The type \mbox{\isa{{\isasymtau}\isactrlsub {\isadigit{1}}\ {\isasymrightharpoonup}\ {\isasymtau}\isactrlsub {\isadigit{2}}}} abbreviates \isa{{\isasymtau}\isactrlsub {\isadigit{1}}\ {\isasymRightarrow}\ {\isasymtau}\isactrlsub {\isadigit{2}}\ option}, i.e.\ partial functions, which we call \emph{maps}.
Maps have a domain and a range:
\begin{trivlist}
\item \isa{\textsf{dom}\ m\ {\isacharequal}\ {\isacharbraceleft}a\ {\isacharbar}\ m\ a\ {\isasymnoteq}\ \textsf{None}{\isacharbraceright}} \qquad \isa{\textsf{ran}\ m\ {\isacharequal}\ {\isacharbraceleft}b\ {\isacharbar}\ {\isasymexists}a{\isachardot}\ m\ a\ {\isacharequal}\ \textsf{Some}\ b{\isacharbraceright}}.
\end{trivlist}

Logical equivalence is written \isa{{\isacharequal}} instead of \isa{{\isasymlongleftrightarrow}}.

\section{Types and Terms}\label{sec:typterm}

A \isa{name} is simply a string. Variables have type \isa{var}; their inner structure is immaterial
for the presentation of the logic.

The logic has three layers: terms are classified by types as usual, but in addition types are
classified by \emph{sorts}. A \isa{sort} is simply a set of class names.
We discuss sorts in detail later.

Types (typically denoted by \isa{T}, \isa{U}, \dots) are defined like this:
\begin{isabelle}%
\isacommand{datatype}\ typ\ {\isacharequal}\ \textsf{Ty}\ name\ {\isacharparenleft}typ\ list{\isacharparenright}\ {\isacharbar}\ \textsf{Tv}\ var\ sort%
\end{isabelle}
where \isa{\textsf{Ty}} \isa{{\isasymkappa}\ {\isacharbrackleft}T\isactrlsub {\isadigit{1}}{\isacharcomma}{\isachardot}{\isachardot}{\isachardot}{\isacharcomma}T\isactrlsub n{\isacharbrackright}} represents the Isabelle type \isa{{\isacharparenleft}T\isactrlsub {\isadigit{1}}{\isacharcomma}{\isasymdots}{\isacharcomma}T\isactrlsub n{\isacharparenright}\ {\isasymkappa}}
and \isa{\textsf{Tv}\ a\ S} represents
a type variable \isa{a} of sort \isa{S} --- sorts are directly attached to type variables.
The notation \isa{T\ {\isasymrightarrow}\ U} is short for \withquotes{\isa{\textsf{Ty}} \isa{{\isachardoublequote}fun{\isachardoublequote}\ {\isacharbrackleft}T{\isacharcomma}U{\isacharbrackright}}, where \isa{{\isachardoublequote}fun{\isachardoublequote}}} is the name
of the function type constructor.

Isabelle's terms are simply typed lambda terms in De Bruijn notation:
\begin{isabelle}%
\isacommand{datatype}\ term\ {\isacharequal}\ \textsf{Ct}\ name\ typ\ {\isacharbar}\ \textsf{Fv}\ var\ typ\ {\isacharbar}\ \textsf{Bv}\ nat\ {\isacharbar}\ \textsf{Abs}\ typ\ term\ {\isacharbar}\ {\isacharparenleft}{\isasymbullet}{\isacharparenright}\ term\ term%
\end{isabelle}
A term (typically \isa{r}, \isa{s}, \isa{t}, \isa{u} \dots) can be a typed constant \isa{\textsf{Ct}\ c\ T} or free variable
\mbox{\isa{\textsf{Fv}\ v\ T}},
a bound variable \isa{\textsf{Bv}\ n} (a De Brujin index), a typed
abstraction \mbox{\isa{\textsf{Abs}\ T\ t}} or an application \isa{t\ {\isasymbullet}\ u}.

The term-has-type proposition has the syntax \isa{Ts\ {\isasymturnstile}\isactrlsub {\isasymtau}\ t\ {\isacharcolon}\ T} where \isa{Ts} is a list of types,
the context for the type of the bound variables.
\begin{center}
\isa{\_\ {\isasymturnstile}\isactrlsub {\isasymtau}\ \textsf{Ct}\ \_\ T\ {\isacharcolon}\ T} \qquad
\isa{\_\ {\isasymturnstile}\isactrlsub {\isasymtau}\ \textsf{Fv}\ \_\ T\ {\isacharcolon}\ T} \qquad
\isa{\mbox{}\inferrule{\mbox{i\ {\isacharless}\ {\isacharbar}Ts{\isacharbar}}}{\mbox{Ts\ {\isasymturnstile}\isactrlsub {\isasymtau}\ \textsf{Bv}\ i\ {\isacharcolon}\ Ts\ {\isacharbang}\ i}}}
\smallskip

\isa{\mbox{}\inferrule{\mbox{T\ {\isacharhash}\ Ts\ {\isasymturnstile}\isactrlsub {\isasymtau}\ t\ {\isacharcolon}\ T{\isacharprime}}}{\mbox{Ts\ {\isasymturnstile}\isactrlsub {\isasymtau}\ \textsf{Abs}\ T\ t\ {\isacharcolon}\ T\ {\isasymrightarrow}\ T{\isacharprime}}}}
\smallskip

\isa{\mbox{}\inferrule{\mbox{Ts\ {\isasymturnstile}\isactrlsub {\isasymtau}\ u\ {\isacharcolon}\ U}\\\ \mbox{Ts\ {\isasymturnstile}\isactrlsub {\isasymtau}\ t\ {\isacharcolon}\ U\ {\isasymrightarrow}\ T}}{\mbox{Ts\ {\isasymturnstile}\isactrlsub {\isasymtau}\ t\ {\isasymbullet}\ u\ {\isacharcolon}\ T}}}
\end{center}
We define \ \isa{{\isasymturnstile}\isactrlsub {\isasymtau}\ t\ {\isacharcolon}\ T\ {\isacharequal}\ {\isacharbrackleft}{\isacharbrackright}\ {\isasymturnstile}\isactrlsub {\isasymtau}\ t\ {\isacharcolon}\ T}.

Function \isa{\textsf{fv}\ {\isacharcolon}{\isacharcolon}\ term\ {\isasymRightarrow}\ {\isacharparenleft}var\ {\isasymtimes}\ typ{\isacharparenright}\ set} collects the free variables in a term. Because bound variables are indices,
\isa{\textsf{fv}\ t} is simply the set of all \isa{{\isacharparenleft}v{\isacharcomma}\ T{\isacharparenright}} such that \isa{\textsf{Fv}\ v\ T} occurs in \isa{t}.
The type is an integral part of a variable.

A \emph{type substitution} is a function \isa{{\isasymrho}} of type \isa{var\ {\isasymRightarrow}\ sort\ {\isasymRightarrow}\ typ}. It assigns a type to each
type variable and sort pair. We write \isa{{\isasymrho}\ {\isachardollar}{\isachardollar}\ T} or \isa{{\isasymrho}\ {\isachardollar}{\isachardollar}\ t} for the overloaded function which applies such a type substitution to
all type variables (and their sort) occurring in a type or term.
The \emph{type instance} relation is defined like this:
\begin{isabelle}%
T\isactrlsub {\isadigit{1}}\ {\isasymlesssim}\ T\isactrlsub {\isadigit{2}}\ {\isacharequal}\ {\isacharparenleft}{\isasymexists}{\isasymrho}{\isachardot}\ {\isasymrho}\ {\isachardollar}{\isachardollar}\ T\isactrlsub {\isadigit{2}}\ {\isacharequal}\ T\isactrlsub {\isadigit{1}}{\isacharparenright}%
\end{isabelle}

We also need to $\beta$-contract a term \isa{\textsf{Abs}\ T\ t\ {\isasymbullet}\ u} to something like
``\isa{t} with \mbox{\isa{\textsf{Bv}\ {\isadigit{0}}}} replaced by \isa{u}''. We define a function \isa{\textsf{subst-bv}}
such that \isa{\textsf{subst-bv}\ u\ t} is that $\beta$-contractum.
The definition of \isa{\textsf{subst-bv}} is shown in the Appendix and can also be found in
the literature (e.g.\ \cite{Nipkow-JAR01}).

In order to abstract over a free (term) variable
there is a function \isa{\textsf{bind-fv}\ {\isacharparenleft}v{\isacharcomma}\ T{\isacharparenright}\ t} that (roughly speaking)
replaces all occurrences of \isa{\textsf{Fv}\ v\ T} in \isa{t} by \isa{\textsf{Bv}\ {\isadigit{0}}}.
Again, see the Appendix for the definition.
This produces (if \isa{\textsf{Fv}\ v\ T} occurs in \isa{t}) a term
with an unbound \isa{\textsf{Bv}\ {\isadigit{0}}}. Function \isa{\textsf{Abs-fv}} binds it with an abstraction:
{\renewcommand{\isasymequiv}{=}%
\begin{isabelle}%
\textsf{Abs-fv}\ v\ T\ t\ {\isasymequiv}\ \textsf{Abs}\ T\ {\isacharparenleft}\textsf{bind-fv}\ {\isacharparenleft}v{\isacharcomma}\ T{\isacharparenright}\ t{\isacharparenright}%
\end{isabelle}}

While this section described the syntax of types and terms, they are not necessarily wellformed and should
be considered pretypes/preterms. The wellformedness checks are described later.

\section{Classes and Sorts}

Isabelle has a built-in system of type classes \cite{Nipkow-LF-91} as in Haskell 98
except that class constraints are directly attached to variable names: our \isa{Tv\ a\ {\isacharbrackleft}C{\isacharcomma}D{\isacharcomma}{\isasymdots}{\isacharbrackright}}
corresponds to Haskell's \texttt{(C a, D a, ...) => ... a ...}.

A \isa{sort} is Isabelle's terminology for a set of (class) names, e.g. \isa{{\isacharbraceleft}C{\isacharcomma}D{\isacharcomma}{\isasymdots}{\isacharbraceright}},
which represent a conjunction of class constraints. In our work, variables \isa{S}, \isa{S{\isacharprime}} etc.\
stand for sorts.

Apart from the usual application in object logics,
type classes also serve an important metalogical purpose: they allow us to restrict, for example,
quantification in object logics to object-level types and rule out meta-level propositions.

Isabelle's type class system was first presented in a programming language context \cite{Nipkow-Snelting,Nipkow-Prehofer-JFP}.
We give the first machine-checked formalization.
The central data structure is a so-called \emph{order-sorted signature}.
Intuitively, it is comprised of a set of class names, a partial subclass ordering on them
and a set of \emph{type constructor signatures}. A type constructor signature \isa{{\isasymkappa}\ {\isacharcolon}{\isacharcolon}\ {\isacharparenleft}S\isactrlsub {\isadigit{1}}{\isacharcomma}\ {\isasymdots}{\isacharcomma}\ S\isactrlsub k{\isacharparenright}\ c}
for a type constructor \isa{{\isasymkappa}} states that applying \isa{{\isasymkappa}} to types \isa{T\isactrlsub {\isadigit{1}}{\isacharcomma}\ {\isasymdots}{\isacharcomma}\ T\isactrlsub k}
such that \isa{T\isactrlsub i} has sort \isa{S\isactrlsub i} (defined below) produces a type of class \isa{c}.
Formally:
\begin{trivlist}
\item
\isakeyword{type\_synonym} \isa{osig} = \noquotes{\isa{{\isachardoublequote}{\isacharparenleft}{\isacharparenleft}name\ {\isasymtimes}\ name{\isacharparenright}\ set\ {\isasymtimes}\ {\isacharparenleft}name\ {\isasymrightharpoonup}\ {\isacharparenleft}class\ {\isasymrightharpoonup}\ sort\ list{\isacharparenright}{\isacharparenright}{\isacharparenright}{\isachardoublequote}}}
\end{trivlist}
To explain this formalization we start from a pair \noquotes{\isa{{\isachardoublequote}{\isacharparenleft}sub{\isacharcomma}tcs{\isacharparenright}\ {\isacharcolon}{\isacharcolon}\ osig{\isachardoublequote}}} and recover the
informal order-sorted signature described above.
The set of classes is simply the \isa{\textsf{Field}} of the \isa{sub} relation.
The \isa{tcs} component represents the set of all type constructor signatures \isa{{\isasymkappa}\ {\isacharcolon}{\isacharcolon}\ {\isacharparenleft}Ss{\isacharparenright}\ c}
(where \isa{Ss} is a list of sorts) such that \isa{tcs\ {\isasymkappa}\ {\isacharequal}\ \textsf{Some}\ dm} and \isa{dm\ c\ {\isacharequal}\ \textsf{Some}\ Ss}.
Representing \isa{{\isasymkappa}\ {\isacharcolon}{\isacharcolon}\ {\isacharparenleft}Ss{\isacharparenright}\ c} as a triple, we define
\begin{trivlist}
\item \isa{TCS\ {\isacharequal}} \isa{{\isacharbraceleft}{\isacharparenleft}{\isasymkappa}{\isacharcomma}\ Ss{\isacharcomma}\ c{\isacharparenright}\ {\isacharbar}\ {\isasymexists}domf{\isachardot}\ tcs\ {\isasymkappa}\ {\isacharequal}\ \textsf{Some}\ domf\ {\isasymand}\ domf\ c\ {\isacharequal}\ \textsf{Some}\ Ss{\isacharbraceright}}
\end{trivlist}
\isa{TCS} is the translation of \isa{tcs}, the data structure close to the implementation,
to an equivalent but more intuitive version \isa{TCS} that is close to the informal presentations
in the literature.

The subclass ordering \isa{sub} can be extended to a subsort ordering as follows:
\begin{isabelle}%
S\isactrlsub {\isadigit{1}}\ {\isasymle}\isactrlbsub sub\isactrlesub \ S\isactrlsub {\isadigit{2}}\ {\isacharequal}\ {\isacharparenleft}{\isasymforall}c\isactrlsub {\isadigit{2}}{\isasymin}S\isactrlsub {\isadigit{2}}{\isachardot}\ {\isasymexists}c\isactrlsub {\isadigit{1}}{\isasymin}S\isactrlsub {\isadigit{1}}{\isachardot}\ c\isactrlsub {\isadigit{1}}\ {\isasymle}\isactrlbsub sub\isactrlesub \ c\isactrlsub {\isadigit{2}}{\isacharparenright}%
\end{isabelle}
The smaller sort needs to subsume all the classes in the larger sort.
In particular \isa{{\isacharbraceleft}c\isactrlsub {\isadigit{1}}{\isacharbraceright}\ {\isasymle}\isactrlbsub sub\isactrlesub \ {\isacharbraceleft}c\isactrlsub {\isadigit{2}}{\isacharbraceright}} iff \isa{{\isacharparenleft}c\isactrlsub {\isadigit{1}}{\isacharcomma}\ c\isactrlsub {\isadigit{2}}{\isacharparenright}\ {\isasymin}\ sub}.

Now we can define a predicate \isa{\textsf{has-sort}} that checks whether, in the context of
some order-sorted signature \isa{{\isacharparenleft}sub{\isacharcomma}tcs{\isacharparenright}}, a type fulfills a given sort constraint:
\begin{center}%
\isa{\mbox{}\inferrule{\mbox{S\ {\isasymle}\isactrlbsub sub\isactrlesub \ S{\isacharprime}}}{\mbox{\textsf{has-sort}\ {\isacharparenleft}sub{\isacharcomma}\ tcs{\isacharparenright}\ {\isacharparenleft}\textsf{Tv}\ a\ S{\isacharparenright}\ S{\isacharprime}}}}
\smallskip

\isa{\mbox{}\inferrule{\mbox{tcs\ {\isasymkappa}\ {\isacharequal}\ \textsf{Some}\ dm}\\\ \mbox{{\isasymforall}c{\isasymin}S{\isachardot}\ {\isasymexists}Ss{\isachardot}\ dm\ c\ {\isacharequal}\ \textsf{Some}\ Ss\ {\isasymand}\ \textsf{list-all2}\ {\isacharparenleft}\textsf{has-sort}\ {\isacharparenleft}sub{\isacharcomma}\ tcs{\isacharparenright}{\isacharparenright}\ Ts\ Ss}}{\mbox{\textsf{has-sort}\ {\isacharparenleft}sub{\isacharcomma}\ tcs{\isacharparenright}\ {\isacharparenleft}\textsf{Ty}\ {\isasymkappa}\ Ts{\isacharparenright}\ S}}}
\end{center}
The rule for type variables uses the subsort relation and is obvious.
A type \isa{{\isacharparenleft}T\isactrlsub {\isadigit{1}}{\isacharcomma}\ {\isasymdots}{\isacharcomma}\ T\isactrlsub n{\isacharparenright}\ {\isasymkappa}} has sort \isa{{\isacharbraceleft}c\isactrlsub {\isadigit{1}}{\isacharcomma}\ {\isasymdots}{\isacharbraceright}} if for every \isa{c\isactrlsub i}
there is a signature \isa{{\isasymkappa}\ {\isacharcolon}{\isacharcolon}\ {\isacharparenleft}S\isactrlsub {\isadigit{1}}{\isacharcomma}\ {\isasymdots}{\isacharcomma}\ S\isactrlsub n{\isacharparenright}\ c\isactrlsub i} and \isa{\textsf{has-sort}\ {\isacharparenleft}sub{\isacharcomma}\ tcs{\isacharparenright}\ T\isactrlsub j\ S\isactrlsub j}
for \isa{j\ {\isacharequal}\ {\isadigit{1}}{\isacharcomma}\ {\isasymdots}{\isacharcomma}\ n}.

We \emph{normalize} a sort by removing ``superfluous" class constraints, i.e.\
retaining only those classes that are not subsumed by other classes.
This gives us unique representatives for sorts which we call \emph{normalized}:
\begin{trivlist}
\item
\isa{\textsf{normalize-sort}\ sub\ S\ {\isacharequal}\ {\isacharbraceleft}c\ {\isasymin}\ S\ {\isacharbar}\ {\isasymnot}\ {\isacharparenleft}{\isasymexists}c{\isacharprime}{\isasymin}S{\isachardot}\ {\isacharparenleft}c{\isacharprime}{\isacharcomma}\ c{\isacharparenright}\ {\isasymin}\ sub\ {\isasymand}\ {\isacharparenleft}c{\isacharcomma}\ c{\isacharprime}{\isacharparenright}\ {\isasymnotin}\ sub{\isacharparenright}{\isacharbraceright}}\\
\isa{\textsf{normalized-sort}\ sub\ S\ {\isacharequal}\ {\isacharparenleft}\textsf{normalize-sort}\ sub\ S\ {\isacharequal}\ S{\isacharparenright}}
\end{trivlist}
We work with normalized sorts because it simplifies the derivation of efficient executable code later on.

Now we can define wellformedness of an \isa{osig}:
\begin{isabelle}%
\textsf{wf-osig}\ {\isacharparenleft}sub{\isacharcomma}\ tcs{\isacharparenright}\ {\isacharequal}\ {\isacharparenleft}\textsf{wf-subclass}\ sub\ {\isasymand}\ \textsf{wf-tcsigs}\ sub\ tcs{\isacharparenright}%
\end{isabelle}
A sublass relation is wellformed if it is a partial order where reflexivity is restricted to
its \isa{\textsf{Field}}.
Wellformedness of type constructor signatures (\isa{\textsf{wf-tcsigs}}) is more complex.
We describe it in terms of \isa{TCS} derived from \isa{tcs} (see above). The conditions are the following:
\begin{itemize}
\item
The following property requires a) that for any \isa{{\isasymkappa}\ {\isacharcolon}{\isacharcolon}\ {\isacharparenleft}{\isachardot}{\isachardot}{\isachardot}{\isacharparenright}c\isactrlsub {\isadigit{1}}} there must be a \mbox{\isa{{\isasymkappa}\ {\isacharcolon}{\isacharcolon}\ {\isacharparenleft}{\isachardot}{\isachardot}{\isachardot}{\isacharparenright}c\isactrlsub {\isadigit{2}}}}
for every superclass \isa{c\isactrlsub {\isadigit{2}}} of \isa{c\isactrlsub {\isadigit{1}}} and b) \emph{coregularity}
which guarantees the existence of principal types \cite{Nipkow-Snelting,types/HaftmannW06}.

\isa{{\isasymforall}{\isacharparenleft}{\isasymkappa}{\isacharcomma}\ Ss\isactrlsub {\isadigit{1}}{\isacharcomma}\ c\isactrlsub {\isadigit{1}}{\isacharparenright}{\isasymin}TCS{\isachardot}\isanewline
\isaindent{\ \ \ }{\isasymforall}c\isactrlsub {\isadigit{2}}{\isachardot}\ {\isacharparenleft}c\isactrlsub {\isadigit{1}}{\isacharcomma}\ c\isactrlsub {\isadigit{2}}{\isacharparenright}\ {\isasymin}\ sub\ {\isasymlongrightarrow}\isanewline
\isaindent{\ \ \ {\isasymforall}c\isactrlsub {\isadigit{2}}{\isachardot}\ }{\isacharparenleft}{\isasymexists}Ss\isactrlsub {\isadigit{2}}{\isachardot}\ {\isacharparenleft}{\isasymkappa}{\isacharcomma}\ Ss\isactrlsub {\isadigit{2}}{\isacharcomma}\ c\isactrlsub {\isadigit{2}}{\isacharparenright}\ {\isasymin}\ TCS\ {\isasymand}\ \textsf{list-all2}\ {\isacharparenleft}{\isasymlambda}S\isactrlsub {\isadigit{1}}\ S\isactrlsub {\isadigit{2}}{\isachardot}\ S\isactrlsub {\isadigit{1}}\ {\isasymle}\isactrlbsub sub\isactrlesub \ S\isactrlsub {\isadigit{2}}{\isacharparenright}\ Ss\isactrlsub {\isadigit{1}}\ Ss\isactrlsub {\isadigit{2}}{\isacharparenright}}

\item A type constructor must always take the same number of argument types:

\isa{{\isasymforall}{\isasymkappa}\ Ss\isactrlsub {\isadigit{1}}\ c\isactrlsub {\isadigit{1}}\ Ss\isactrlsub {\isadigit{2}}\ c\isactrlsub {\isadigit{2}}{\isachardot}\isanewline
\isaindent{\ \ \ }{\isacharparenleft}{\isasymkappa}{\isacharcomma}\ Ss\isactrlsub {\isadigit{1}}{\isacharcomma}\ c\isactrlsub {\isadigit{1}}{\isacharparenright}\ {\isasymin}\ TCS\ {\isasymand}\ {\isacharparenleft}{\isasymkappa}{\isacharcomma}\ Ss\isactrlsub {\isadigit{2}}{\isacharcomma}\ c\isactrlsub {\isadigit{2}}{\isacharparenright}\ {\isasymin}\ TCS\ {\isasymlongrightarrow}\ {\isacharbar}Ss\isactrlsub {\isadigit{1}}{\isacharbar}\ {\isacharequal}\ {\isacharbar}Ss\isactrlsub {\isadigit{2}}{\isacharbar}}

\item Sorts must be normalized and must exists in \isa{sub}:

\isa{{\isasymforall}{\isacharparenleft}{\isasymkappa}{\isacharcomma}\ Ss{\isacharcomma}\ c{\isacharparenright}{\isasymin}TCS{\isachardot}\ {\isasymforall}S{\isasymin}\textsf{set}\ Ss{\isachardot}\ \textsf{wf-sort}\ sub\ S}

where \isa{\textsf{wf-sort}\ sub\ S\ {\isacharequal}\ {\isacharparenleft}\textsf{normalized-sort}\ sub\ S\ {\isasymand}\ S\ {\isasymsubseteq}\ \textsf{Field}\ sub{\isacharparenright}}
\end{itemize}
These conditions are used in a number of places to show that the type system is well behaved.
For example, \isa{\textsf{has-sort}} is upward closed:
\begin{trivlist}
\item
\isa{\textsf{wf-osig}\ {\isacharparenleft}sub{\isacharcomma}\ tcs{\isacharparenright}} \isa{{\isasymand}} \isa{\textsf{has-sort}\ {\isacharparenleft}sub{\isacharcomma}\ tcs{\isacharparenright}\ T\ S} \isa{{\isasymand}} \isa{S\ {\isasymle}\isactrlbsub sub\isactrlesub \ S{\isacharprime}}\\
\isa{{\isasymlongrightarrow}} \isa{\textsf{has-sort}\ {\isacharparenleft}sub{\isacharcomma}\ tcs{\isacharparenright}\ T\ S{\isacharprime}}
\end{trivlist}

\section{Signatures}

A \emph{signature} consist of a map from constant names to their (most general) types,
a map from type constructor names to their arities, and an order-sorted signature:
\begin{trivlist}
\item\isakeyword{type\_synonym} \isa{signature} = \noquotes{\isa{{\isachardoublequote}{\isacharparenleft}name\ {\isasymrightharpoonup}\ typ{\isacharparenright}\ {\isasymtimes}\ {\isacharparenleft}name\ {\isasymrightharpoonup}\ nat{\isacharparenright}\ {\isasymtimes}\ osig{\isachardoublequote}}}
\end{trivlist}
The three projection functions are called \isa{\textsf{const-type}}, \isa{\textsf{type-arity}}
and \isa{\textsf{osig}}.
We now define a number of wellformedness checks w.r.t.\ a signature \isa{{\isasymSigma}}.
We start with wellformedness of types, which is pretty obvious:
\begin{center}
\isa{\mbox{}\inferrule{\mbox{\textsf{type-arity}\ {\isasymSigma}\ {\isasymkappa}\ {\isacharequal}\ \textsf{Some}\ {\isacharbar}Ts{\isacharbar}}\\\ \mbox{{\isasymforall}T{\isasymin}\textsf{set}\ Ts{\isachardot}\ \textsf{wf-type}\ {\isasymSigma}\ T}}{\mbox{\textsf{wf-type}\ {\isasymSigma}\ {\isacharparenleft}\textsf{Ty}\ {\isasymkappa}\ Ts{\isacharparenright}}}}
\smallskip

\isa{\mbox{}\inferrule{\mbox{\textsf{wf-sort}\ {\isacharparenleft}\textsf{subclass}\ {\isacharparenleft}\textsf{osig}\ {\isasymSigma}{\isacharparenright}{\isacharparenright}\ S}}{\mbox{\textsf{wf-type}\ {\isasymSigma}\ {\isacharparenleft}\textsf{Tv}\ a\ S{\isacharparenright}}}}
\end{center}

Wellformedness of a term essentially just says that all types in the term are wellformed
and that the type \isa{T{\isacharprime}} of a constant in the term must be an instance of the type \isa{T} of that constant
in the signature: \isa{T{\isacharprime}\ {\isasymlesssim}\ T}.
\begin{center}
\isa{\mbox{}\inferrule{\mbox{\textsf{wf-type}\ {\isasymSigma}\ T}}{\mbox{\textsf{wf-term}\ {\isasymSigma}\ {\isacharparenleft}\textsf{Fv}\ v\ T{\isacharparenright}}}}
\qquad
\isa{\textsf{wf-term}\ {\isasymSigma}\ {\isacharparenleft}\textsf{Bv}\ n{\isacharparenright}}
\smallskip

\isa{\mbox{}\inferrule{\mbox{\textsf{const-type}\ {\isasymSigma}\ s\ {\isacharequal}\ \textsf{Some}\ T}\\\ \mbox{\textsf{wf-type}\ {\isasymSigma}\ T{\isacharprime}}\\\ \mbox{T{\isacharprime}\ {\isasymlesssim}\ T}}{\mbox{\textsf{wf-term}\ {\isasymSigma}\ {\isacharparenleft}\textsf{Ct}\ s\ T{\isacharprime}{\isacharparenright}}}}
\smallskip

\isa{\mbox{}\inferrule{\mbox{\textsf{wf-term}\ {\isasymSigma}\ t}\\\ \mbox{\textsf{wf-term}\ {\isasymSigma}\ u}}{\mbox{\textsf{wf-term}\ {\isasymSigma}\ {\isacharparenleft}t\ {\isasymbullet}\ u{\isacharparenright}}}}
\smallskip

\isa{\mbox{}\inferrule{\mbox{\textsf{wf-type}\ {\isasymSigma}\ T}\\\ \mbox{\textsf{wf-term}\ {\isasymSigma}\ t}}{\mbox{\textsf{wf-term}\ {\isasymSigma}\ {\isacharparenleft}\textsf{Abs}\ T\ t{\isacharparenright}}}}
\end{center}
These rules only check whether a term conforms to a signature, not that the contained types are consistent.
Combining wellformedness and \isa{{\isasymturnstile}\isactrlsub {\isasymtau}} yields welltypedness of a term:

\begin{isabelle}%
\textsf{wt-term}\ {\isasymSigma}\ t\ {\isacharequal}\ {\isacharparenleft}\textsf{wf-term}\ {\isasymSigma}\ t\ {\isasymand}\ {\isacharparenleft}{\isasymexists}T{\isachardot}\ {\isasymturnstile}\isactrlsub {\isasymtau}\ t\ {\isacharcolon}\ T{\isacharparenright}{\isacharparenright}%
\end{isabelle}

Wellformedness of a signature \isa{{\isasymSigma}\ {\isacharequal}\ {\isacharparenleft}ctf{\isacharcomma}\ arf{\isacharcomma}\ oss{\isacharparenright}} where \isa{oss\ {\isacharequal}\ {\isacharparenleft}sub{\isacharcomma}\ tcs{\isacharparenright}} is defined
as follows:
\begin{isabelle}%
\textsf{wf-sig}\ {\isasymSigma}\ {\isacharequal}\isanewline
{\isacharparenleft}{\isacharparenleft}{\isasymforall}T{\isasymin}\textsf{ran}\ ctf{\isachardot}\ \textsf{wf-type}\ {\isasymSigma}\ T{\isacharparenright}\ {\isasymand}\ \textsf{wf-osig}\ oss\ {\isasymand}\ \textsf{dom}\ tcs\ {\isacharequal}\ \textsf{dom}\ arf\ {\isasymand}\isanewline
\isaindent{{\isacharparenleft}}{\isacharparenleft}{\isasymforall}{\isasymkappa}\ dm{\isachardot}\ tcs\ {\isasymkappa}\ {\isacharequal}\ \textsf{Some}\ dm\ {\isasymlongrightarrow}\ {\isacharparenleft}{\isasymforall}Ss{\isasymin}\textsf{ran}\ dm{\isachardot}\ arf\ {\isasymkappa}\ {\isacharequal}\ \textsf{Some}\ {\isacharbar}Ss{\isacharbar}{\isacharparenright}{\isacharparenright}{\isacharparenright}%
\end{isabelle}
In words: all types in \isa{ctf} are wellformed,
 \isa{oss} is wellformed,
 the type constructors in \isa{tcs} are exactly those that have an arity in \isa{arf},
 for every type constructor signature \isa{{\isacharparenleft}{\isasymkappa}{\isacharcomma}\ Ss{\isacharcomma}\ \_{\isacharparenright}} in \isa{tcs}, 
 \isa{{\isasymkappa}} has arity \isa{{\isacharbar}Ss{\isacharbar}}.

\section{Logic}

Isabelle's metalogic \M\ is an extension of the logic described by Paulson \cite{Paulson-JAR-89}.
It is a fragment of intuitionistic higher-order logic. The basic types and connectives of \M\ are the following:
\begin{center}
\begin{tabular}{|l|l|c|}
\hline
Concept & Representation & Abbreviation\\ \hline
Type of propositions & \withquotes{\isa{\textsf{Ty}} \isa{{\isachardoublequote}prop{\isachardoublequote}\ {\isacharbrackleft}{\isacharbrackright}}} & \isa{\textsf{prop}}\\
Implication & \isa{\textsf{Ct}} \withquotes{\isa{{\isachardoublequote}imp{\isachardoublequote}}} \isa{{\isacharparenleft}}\isa{\textsf{prop}\ {\isasymrightarrow}\ \textsf{prop}\ {\isasymrightarrow}\ \textsf{prop}}\isa{{\isacharparenright}} & \isa{{\isasymLongrightarrow}} \\
Universal quantifier & \isa{\textsf{Ct}} \withquotes{\isa{{\isachardoublequote}all{\isachardoublequote}}} \isa{{\isacharparenleft}}\isa{{\isacharparenleft}T\ {\isasymrightarrow}\ \textsf{prop}{\isacharparenright}\ {\isasymrightarrow}\ \textsf{prop}}\isa{{\isacharparenright}} & \isa{{\isasymAnd}\isactrlsub T}\\
Equality & \isa{\textsf{Ct}} \withquotes{\isa{{\isachardoublequote}eq{\isachardoublequote}}} \isa{{\isacharparenleft}}\isa{T\ {\isasymrightarrow}\ T\ {\isasymrightarrow}\ \textsf{prop}}\isa{{\isacharparenright}} & \isa{{\isasymequiv}\isactrlsub T} \\ \hline
\end{tabular}
\end{center}
The type subscripts of \isa{{\isasymAnd}} and \isa{{\isasymequiv}} are dropped in the text if they can be inferred.

Readers familiar with Isabelle syntax must keep in mind that for readability we use the symbols \isa{{\isasymAnd}}, \isa{{\isasymLongrightarrow}} and \isa{{\isasymequiv}}
for the \emph{encodings} of the respective symbols in Isabelle's metalogic. We avoid the corresponding
metalogical constants completely in favour of HOL's \isa{{\isasymforall}}, \isa{{\isasymlongrightarrow}}, \isa{{\isacharequal}} and inference rule notation.

The provability judgment of \M\ is of the form \isa{{\isasymTheta}{\isacharcomma}{\isasymGamma}\ {\isasymturnstile}\ t} where \isa{{\isasymTheta}} is a theory,
\isa{{\isasymGamma}} (the hypotheses) is a set of terms of type \isa{\textsf{prop}} and \isa{t} a term of type \isa{\textsf{prop}}.

A \emph{theory} is a pair of a signature and a set of axioms:
\begin{trivlist}
\item \isakeyword{type\_synonym} \isa{theory} = \noquotes{\isa{{\isachardoublequote}signature\ {\isasymtimes}\ term\ set{\isachardoublequote}}}
\end{trivlist}
The projection functions are \isa{\textsf{sig}} and \isa{\textsf{axioms}}.
We extend the notion of wellformedness from signatures to theories:
\begin{isabelle}%
\textsf{wf-theory}\ {\isacharparenleft}{\isasymSigma}{\isacharcomma}\ axs{\isacharparenright}\ {\isacharequal}\isanewline
{\isacharparenleft}\textsf{wf-sig}\ {\isasymSigma}\ {\isasymand}\ {\isacharparenleft}{\isasymforall}p{\isasymin}axs{\isachardot}\ \textsf{wt-term}\ {\isasymSigma}\ p\ {\isasymand}\ {\isasymturnstile}\isactrlsub {\isasymtau}\ p\ {\isacharcolon}\ \textsf{prop}{\isacharparenright}\ {\isasymand}\ \textsf{is-std-sig}\ {\isasymSigma}\ {\isasymand}\ \textsf{eq-axs}\ {\isasymsubseteq}\ axs{\isacharparenright}%
\end{isabelle}
The first two conjuncts need no explanation.
Predicate \isa{\textsf{is-std-sig}} (not shown) requires the signature to have certain minimal content:
the basic types (\isa{{\isasymrightarrow}}, \isa{\textsf{prop}}) and constants (\isa{{\isasymequiv}}, \isa{{\isasymAnd}}, \isa{{\isasymLongrightarrow}})
of \M\ and the additional types and constants for type class reasoning
from Section~\ref{sec:TypeClassReasoning}.
Our theories also need to contain a minimal set of axioms.
The set \isa{\textsf{eq-axs}} is an axiomatic basis for equality reasoning and
will be explained in Section~\ref{sec:Equality}.

We will now discuss the inference system in three steps:
the basic inference rules, equality and type class reasoning.

\subsection{Basic Inference Rules}

The \emph{axiom rule} states that wellformed type-instances of axioms are provable:
\begin{center}
\isa{\mbox{}\inferrule{\mbox{\textsf{wf-theory}\ {\isasymTheta}}\\\ \mbox{t\ {\isasymin}\ \textsf{axioms}\ {\isasymTheta}}\\\ \mbox{\textsf{wf-inst}\ {\isasymTheta}\ {\isasymrho}}}{\mbox{{\isasymTheta}{\isacharcomma}{\isasymGamma}\ {\isasymturnstile}\ {\isasymrho}\ {\isachardollar}{\isachardollar}\ t}}}
\end{center}
where \isa{{\isasymrho}\ {\isacharcolon}{\isacharcolon}} \isa{var\ {\isasymRightarrow}\ sort\ {\isasymRightarrow}\ typ} is a type substitution and \isa{{\isachardollar}{\isachardollar}}
denotes its application (see Section~\ref{sec:typterm}).
The types substituted into the type variables need to be wellformed and conform to the 
sort constraint of the type variable: 
\begin{isabelle}%
\textsf{wf-inst}\ {\isacharparenleft}{\isasymSigma}{\isacharcomma}\ axs{\isacharparenright}\ {\isasymrho}\ {\isacharequal}\isanewline
{\isacharparenleft}{\isasymforall}v\ S{\isachardot}\ {\isasymrho}\ v\ S\ {\isasymnoteq}\ \textsf{Tv}\ v\ S\ {\isasymlongrightarrow}\ \textsf{has-sort}\ {\isacharparenleft}\textsf{osig}\ {\isasymSigma}{\isacharparenright}\ {\isacharparenleft}{\isasymrho}\ v\ S{\isacharparenright}\ S\ {\isasymand}\ \textsf{wf-type}\ {\isasymSigma}\ {\isacharparenleft}{\isasymrho}\ v\ S{\isacharparenright}{\isacharparenright}%
\end{isabelle}
The conjunction only needs to hold if \isa{{\isasymrho}} actually changes something, i.e.\
if \isa{{\isasymrho}\ v\ S\ {\isasymnoteq}\ \textsf{Tv}\ v\ S}. This condition is not superfluous because
otherwise \isa{\textsf{has-sort}\ oss\ {\isacharparenleft}\textsf{Tv}\ v\ S{\isacharparenright}\ S} and \isa{\textsf{wf-type}\ {\isasymSigma}\ {\isacharparenleft}\textsf{Tv}\ v\ S{\isacharparenright}}
only hold if \isa{S} is wellformed w.r.t\ \isa{{\isasymSigma}}.

Note that there are no extra rules for general instantiation of type or term variables.
Type variables can only be instantiated in the axioms. Term instantiation can be performed using the 
forall introduction and elimination rules.

The \emph{assumption rule} allows us to prove terms already in the hypotheses:
\begin{center}
\isa{\mbox{}\inferrule{\mbox{\textsf{wf-term}\ {\isacharparenleft}\textsf{sig}\ {\isasymTheta}{\isacharparenright}\ t}\\\ \mbox{{\isasymturnstile}\isactrlsub {\isasymtau}\ t\ {\isacharcolon}\ \textsf{prop}}\\\ \mbox{t\ {\isasymin}\ {\isasymGamma}}}{\mbox{{\isasymTheta}{\isacharcomma}{\isasymGamma}\ {\isasymturnstile}\ t}}}
\end{center}

Both \isa{{\isasymAnd}} and \isa{{\isasymLongrightarrow}} are characterized by introduction and elimination rules:
\begin{center}
\isa{\mbox{}\inferrule{\mbox{\textsf{wf-theory}\ {\isasymTheta}}\\\ \mbox{{\isasymTheta}{\isacharcomma}{\isasymGamma}\ {\isasymturnstile}\ t}\\\ \mbox{{\isacharparenleft}x{\isacharcomma}\ T{\isacharparenright}\ {\isasymnotin}\ \textsf{FV}\ {\isasymGamma}}\\\ \mbox{\textsf{wf-type}\ {\isacharparenleft}\textsf{sig}\ {\isasymTheta}{\isacharparenright}\ T}}{\mbox{{\isasymTheta}{\isacharcomma}{\isasymGamma}\ {\isasymturnstile}\ {\isasymAnd}\isactrlsub T\ {\isacharparenleft}\textsf{Abs-fv}\ x\ T\ t{\isacharparenright}}}}
\smallskip

\isa{\mbox{}\inferrule{\mbox{{\isasymTheta}{\isacharcomma}{\isasymGamma}\ {\isasymturnstile}\ {\isasymAnd}\isactrlsub T\ {\isacharparenleft}\textsf{Abs}\ T\ t{\isacharparenright}}\\\ \mbox{{\isasymturnstile}\isactrlsub {\isasymtau}\ u\ {\isacharcolon}\ T}\\\ \mbox{\textsf{wf-term}\ {\isacharparenleft}\textsf{sig}\ {\isasymTheta}{\isacharparenright}\ u}}{\mbox{{\isasymTheta}{\isacharcomma}{\isasymGamma}\ {\isasymturnstile}\ \textsf{subst-bv}\ u\ t}}}
\smallskip

\isa{\mbox{}\inferrule{\mbox{\textsf{wf-theory}\ {\isasymTheta}}\\\ \mbox{{\isasymTheta}{\isacharcomma}{\isasymGamma}\ {\isasymturnstile}\ u}\\\ \mbox{\textsf{wf-term}\ {\isacharparenleft}\textsf{sig}\ {\isasymTheta}{\isacharparenright}\ t}\\\ \mbox{{\isasymturnstile}\isactrlsub {\isasymtau}\ t\ {\isacharcolon}\ \textsf{prop}}}{\mbox{{\isasymTheta}{\isacharcomma}{\isasymGamma}\ {\isacharminus}\ {\isacharbraceleft}t{\isacharbraceright}\ {\isasymturnstile}\ t\ {\isasymLongrightarrow}\ u}}}
\smallskip

\isa{\mbox{}\inferrule{\mbox{{\isasymTheta}{\isacharcomma}{\isasymGamma}\isactrlsub {\isadigit{1}}\ {\isasymturnstile}\ t\ {\isasymLongrightarrow}\ u}\\\ \mbox{{\isasymTheta}{\isacharcomma}{\isasymGamma}\isactrlsub {\isadigit{2}}\ {\isasymturnstile}\ t}}{\mbox{{\isasymTheta}{\isacharcomma}{\isasymGamma}\isactrlsub {\isadigit{1}}\ {\isasymunion}\ {\isasymGamma}\isactrlsub {\isadigit{2}}\ {\isasymturnstile}\ u}}}
\end{center}
where \isa{\textsf{FV}\ {\isasymGamma}\ {\isacharequal}\ {\isacharparenleft}{\isasymUnion}\isactrlbsub t{\isasymin}{\isasymGamma}\isactrlesub \ \textsf{fv}\ t{\isacharparenright}}.

\subsection{Equality}
\label{sec:Equality}

Most rules about equality are not part of the inference system but are axioms
(the set \isa{\textsf{eq-axs}} mentioned above). Consequences are obtained via the axiom rule.

The first three axioms express that \isa{{\isasymequiv}} is reflexive, symmetric and transitive:

\begin{center}
\isa{x\ {\isasymequiv}\ x} \qquad
\isa{x\ {\isasymequiv}\ y\ {\isasymLongrightarrow}\ y\ {\isasymequiv}\ x} \qquad
\isa{x\ {\isasymequiv}\ y\ {\isasymLongrightarrow}\ y\ {\isasymequiv}\ z\ {\isasymLongrightarrow}\ x\ {\isasymequiv}\ z} \qquad
\end{center}

The next two axioms express that terms of type \isa{\textsf{prop}} (\isa{A} and \isa{B})
are equal iff they are logically equivalent:
\begin{center}
\isa{A\ {\isasymequiv}\ B\ {\isasymLongrightarrow}\ A\ {\isasymLongrightarrow}\ B} \qquad
\isa{{\isacharparenleft}A\ {\isasymLongrightarrow}\ B{\isacharparenright}\ {\isasymLongrightarrow}\ {\isacharparenleft}B\ {\isasymLongrightarrow}\ A{\isacharparenright}\ {\isasymLongrightarrow}\ A\ {\isasymequiv}\ B} \smallskip
\end{center}

The last equality axioms are congruence rules for application and abstraction:
\begin{center}
\isa{f\ {\isasymequiv}\ g\ {\isasymLongrightarrow}\ x\ {\isasymequiv}\ y\ {\isasymLongrightarrow}\ {\isacharparenleft}f\ {\isasymbullet}\ x{\isacharparenright}\ {\isasymequiv}\ {\isacharparenleft}g\ {\isasymbullet}\ y{\isacharparenright}} \smallskip

\isa{{\isasymAnd}} \isa{{\isacharparenleft}\textsf{Abs}\ T\ {\isacharparenleft}{\isacharparenleft}f\ {\isasymbullet}\ \textsf{Bv}\ {\isadigit{0}}{\isacharparenright}\ {\isasymequiv}\ {\isacharparenleft}g\ {\isasymbullet}\ \textsf{Bv}\ {\isadigit{0}}{\isacharparenright}{\isacharparenright}{\isacharparenright}} \isa{{\isasymLongrightarrow}} \isa{\textsf{Abs}\ T\ {\isacharparenleft}f\ {\isasymbullet}\ \textsf{Bv}\ {\isadigit{0}}{\isacharparenright}\ {\isasymequiv}\ \textsf{Abs}\ T\ {\isacharparenleft}g\ {\isasymbullet}\ \textsf{Bv}\ {\isadigit{0}}{\isacharparenright}}
\end{center}
Paulson \cite{Paulson-JAR-89} gives a slightly different congruence rule for abstraction, which allows
to abstract over an arbitrary, free \isa{x} in \isa{f{\isacharcomma}g}. We are able to derive this rule in our inference system. 

Finally there are the lambda calculus rules. There is no need for \isa{{\isasymalpha}} conversion because
\isa{{\isasymalpha}}-equivalent terms are already identical thanks to the De Brujin indices for bound variables.
For \isa{{\isasymbeta}} and \isa{{\isasymeta}} conversion the following rules are added.
In contrast to the rest of this subsection, these are not expressed as axioms.
\begin{center}
\isa{\mbox{}\inferrule{\mbox{\textsf{wf-theory}\ {\isasymTheta}}\\\ \mbox{\textsf{wt-term}\ {\isacharparenleft}\textsf{sig}\ {\isasymTheta}{\isacharparenright}\ {\isacharparenleft}\textsf{Abs}\ T\ t{\isacharparenright}}\\\ \mbox{\textsf{wf-term}\ {\isacharparenleft}\textsf{sig}\ {\isasymTheta}{\isacharparenright}\ u}\\\ \mbox{{\isasymturnstile}\isactrlsub {\isasymtau}\ u\ {\isacharcolon}\ T}}{\mbox{{\isasymTheta}{\isacharcomma}{\isasymGamma}\ {\isasymturnstile}\ {\isacharparenleft}\textsf{Abs}\ T\ t\ {\isasymbullet}\ u{\isacharparenright}\ {\isasymequiv}\ \textsf{subst-bv}\ u\ t}}}($\beta$) \smallskip

\isa{\mbox{}\inferrule{\mbox{\textsf{wf-theory}\ {\isasymTheta}}\\\ \mbox{\textsf{wf-term}\ {\isacharparenleft}\textsf{sig}\ {\isasymTheta}{\isacharparenright}\ t}\\\ \mbox{{\isasymturnstile}\isactrlsub {\isasymtau}\ t\ {\isacharcolon}\ T\ {\isasymrightarrow}\ T{\isacharprime}}}{\mbox{{\isasymTheta}{\isacharcomma}{\isasymGamma}\ {\isasymturnstile}\ \textsf{Abs}\ T\ {\isacharparenleft}t\ {\isasymbullet}\ \textsf{Bv}\ {\isadigit{0}}{\isacharparenright}\ {\isasymequiv}\ t}}}($\eta$)
\end{center}
Rule ($\beta$) uses the substitution function \isa{\textsf{subst-bv}} as explained in
Section~\ref{sec:typterm} (and defined in the Appendix).

Rule ($\eta$) requires a few words of explanation. We do not explicitly
require that \isa{t} does not contain \isa{\textsf{Bv}\ {\isadigit{0}}}. This is already a consequence of the precondition
that \isa{{\isasymturnstile}\isactrlsub {\isasymtau}\ t\ {\isacharcolon}\ T\ {\isasymrightarrow}\ T{\isacharprime}}: it implies that \isa{t} is closed. For that reason it is
perfectly unproblematic to remove the abstraction above \isa{t}.

\subsection{Type Class Reasoning}
\label{sec:TypeClassReasoning}

Wenzel \cite{tphol/Wenzel97} encoded class constraints of the form ``type \isa{T} has class \isa{c}''
in the term language as follows.
There is a unary type constructor named \isa{{\isachardoublequote}itself{\isachardoublequote}}
and \isa{T\ \textsf{itself}} abbreviates \withquotes{\isa{\textsf{Ty}} \isa{{\isachardoublequote}itself{\isachardoublequote}\ {\isacharbrackleft}T{\isacharbrackright}}}.
The notation \isa{TYPE\isactrlbsub T\ \textsf{itself}\isactrlesub } is short
for \mbox{\isa{\textsf{Ct}} \withquotes{\isa{{\isachardoublequote}type{\isachardoublequote}\ {\isacharparenleft}}}\isa{T\ \textsf{itself}}\isa{{\isacharparenright}}} where 
\isa{{\isachardoublequote}type{\isachardoublequote}} is the name of a new uninterpreted constant.
You should view \isa{TYPE\isactrlbsub T\ \textsf{itself}\isactrlesub } as the term-level representation of type \isa{T}.

Next we represent the predicate ``is of class \isa{c}'' on the term level.
For this we define some fixed injective mapping \isa{\textsf{const-of-class}} from class to constant names.
For each new class \isa{c} a new constant \isa{\textsf{const-of-class}\ c} of type \isa{T\ \textsf{itself}\ {\isasymrightarrow}\ \textsf{prop}}
is added.
The term \isa{\textsf{Ct}\ {\isacharparenleft}\textsf{const-of-class}\ c{\isacharparenright}\ {\isacharparenleft}T\ \textsf{itself}\ {\isasymrightarrow}\ \textsf{prop}{\isacharparenright}\ {\isasymbullet}\ TYPE\isactrlbsub T\ \textsf{itself}\isactrlesub } represents
the statement ``type \isa{T} has class \isa{c}''. This is the inference rule deriving such propositions:
\begin{center}
\isa{\mbox{}\inferrule{\mbox{\textsf{wf-theory}\ {\isasymTheta}}\\\ \mbox{\textsf{const-type}\ {\isacharparenleft}\textsf{sig}\ {\isasymTheta}{\isacharparenright}\ {\isacharparenleft}\textsf{const-of-class}\ C{\isacharparenright}\ {\isacharequal}\ \textsf{Some}\ {\isacharparenleft}{\isacharprime}a\ \textsf{itself}\ {\isasymrightarrow}\ \textsf{prop}{\isacharparenright}}\\\ \mbox{\textsf{wf-type}\ {\isacharparenleft}\textsf{sig}\ {\isasymTheta}{\isacharparenright}\ T}\\\ \mbox{\textsf{has-sort}\ {\isacharparenleft}\textsf{osig}\ {\isacharparenleft}\textsf{sig}\ {\isasymTheta}{\isacharparenright}{\isacharparenright}\ T\ {\isacharbraceleft}C{\isacharbraceright}}}{\mbox{{\isasymTheta}{\isacharcomma}{\isasymGamma}\ {\isasymturnstile}\ \textsf{Ct}\ {\isacharparenleft}\textsf{const-of-class}\ C{\isacharparenright}\ {\isacharparenleft}T\ \textsf{itself}\ {\isasymrightarrow}\ \textsf{prop}{\isacharparenright}\ {\isasymbullet}\ TYPE\isactrlbsub T\ \textsf{itself}\isactrlesub }}}
\end{center}
This is how the \isa{\textsf{has-sort}} inference system is integrated into the logic.
\bigskip

This concludes the presentation of \M. We have shown some minimal sanity properties,
incl.\ that all provable terms
are of type \isa{\textsf{prop}} and wellformed:
\begin{theorem}
\isa{{\isasymTheta}{\isacharcomma}{\isasymGamma}\ {\isasymturnstile}\ t\ {\isasymlongrightarrow}\ {\isasymturnstile}\isactrlsub {\isasymtau}\ t\ {\isacharcolon}\ \textsf{prop}\ {\isasymand}\ \textsf{wf-term}\ {\isacharparenleft}\textsf{sig}\ {\isasymTheta}{\isacharparenright}\ t}
\end{theorem}

The attentive reader will have noticed that we do not require unused hypotheses in \isa{{\isasymGamma}} to be wellformed and of type \isa{\textsf{prop}}.
Similarly, we only require \isa{\textsf{wf-theory}\ {\isasymTheta}} in rules that need it to preserve
wellformedness of the terms and types involved. To restrict to wellformed theories and hypotheses
we define a top-level provability judgment that requires wellformedness: 
\begin{isabelle}%
{\isasymTheta}{\isacharcomma}{\isasymGamma}\ {\isasymtturnstile}\ t\ {\isacharequal}\ {\isacharparenleft}\textsf{wf-theory}\ {\isasymTheta}\ {\isasymand}\ {\isacharparenleft}{\isasymforall}h{\isasymin}{\isasymGamma}{\isachardot}\ \textsf{wf-term}\ {\isacharparenleft}\textsf{sig}\ {\isasymTheta}{\isacharparenright}\ h\ {\isasymand}\ {\isasymturnstile}\isactrlsub {\isasymtau}\ h\ {\isacharcolon}\ \textsf{prop}{\isacharparenright}\ {\isasymand}\ {\isasymTheta}{\isacharcomma}{\isasymGamma}\ {\isasymturnstile}\ t{\isacharparenright}%
\end{isabelle}

\section{Proof Terms and Checker}

Berghofer and Nipkow \cite{BerghoferN-TPHOLs00} added proof terms to Isabelle. We present an executable
checker for these proof terms that is proved sound w.r.t.\ the above formalization of the metalogic.
Berghofer and Nipkow also developed a proof checker but it was unverified and
checked the generated proof terms by feeding them back through Isabelle's unverified inference kernel.

It is crucial to realize that all we need to know about the  proof term checker is the soundness theorem below.
The internals are, from a soundness perspective, irrelevant, which is why we can get away with sketching
them informally. This is in contrast to the logic itself, which acts like a specification, which is why
we presented it in detail.

This is our data type of proof terms:
\begin{isabelle}%
\isacommand{datatype}\ proofterm\ {\isacharequal}\ \textsf{PAxm}\ term\ {\isacharparenleft}{\isacharparenleft}{\isacharparenleft}var\ {\isasymtimes}\ sort{\isacharparenright}\ {\isasymtimes}\ typ{\isacharparenright}\ list{\isacharparenright}\ {\isacharbar}\ \textsf{PBound}\ nat\isanewline
\isaindent{\ \ }{\isacharbar}\ \textsf{Abst}\ typ\ proofterm\ {\isacharbar}\ \textsf{AbsP}\ term\ proofterm\ {\isacharbar}\ \textsf{Appt}\ proofterm\ term\isanewline
\isaindent{\ \ }{\isacharbar}\ \textsf{AppP}\ proofterm\ proofterm\ {\isacharbar}\ \textsf{OfClass}\ typ\ name\ {\isacharbar}\ \textsf{Hyp}\ term%
\end{isabelle}
These proof terms are not designed to record proofs in our inference system, but
to mirror the proof terms generated by Isabelle. Nevertheless, the constructors of our proof terms
correspond roughly to the rules of the inference system.
\isa{\textsf{PAxm}} contains an axiom and a type substitution. This substitution is encoded as an association
list instead of a function.
\isa{\textsf{AbsP}} and \isa{\textsf{Abst}} correspond to introduction of \isa{{\isasymLongrightarrow}} and \isa{{\isasymAnd}},
\isa{\textsf{AppP}} and \isa{\textsf{Appt}} correspond to the respective eliminations.
\isa{\textsf{Hyp}} and \isa{\textsf{PBound}} relate to the assumption rule, where \isa{\textsf{Hyp}} refers to a free assumption
while \isa{\textsf{PBound}} contains a De Brujin index referring to an assumption added during the proof by an 
\isa{\textsf{AbsP}} constructor. \isa{\textsf{OfClass}} denotes a proof that a type belongs to a given type class. 

Isabelle looks at terms modulo \isa{{\isasymalpha}{\isasymbeta}{\isasymeta}}-equivalence and therefore does not save \isa{{\isasymbeta}} or \isa{{\isasymeta}} steps, 
while they are explicit steps in our inference system.
Therefore we have no constructors corresponding to the ($\beta$) and ($\eta)$ rules. 
The remaining equality axioms are naturally handled by the \isa{\textsf{PAxm}} constructor.

In the rest of the section we discuss how to derive an executable proof checker.
Executability means that the checker is defined as a set of recursive functions
that Isabelle's code generator can translate into one of a number of target languages,
in particular its implementation language SML \cite{BerghoferN-TYPES00,HaftmannN-FLOPS2010,HaftmannKKN-ITP13}.

Because of the approximate correspondence between proof term constructors and inference rules,
implementing the proof checker largely amounts to providing executable versions of each
inference rule, as in LCF: each rule becomes a function that checks the side conditions, and if they are true,
computes the conclusion from the premises given as arguments. The overall checker is a function
\begin{trivlist}
\item \isa{\textsf{replay}} \isa{{\isacharcolon}{\isacharcolon}} \isa{theory\ {\isasymRightarrow}\ proofterm\ {\isasymRightarrow}\ term\ option}
\end{trivlist}
In particular we need to make the inductive wellformedness checks for sorts, types and terms,
signatures and theories executable. 
Mostly, this amounts to providing recursive versions of inductive definitions
and proving them equivalent.

We now discuss some of the more difficult implementation steps.
To model Isabelle's view of terms modulo \isa{{\isasymalpha}{\isasymbeta}{\isasymeta}}-equivalence, we \isa{{\isasymbeta}{\isasymeta}} normalize our terms
(\isa{{\isasymalpha}}-equivalence is for free thanks to De Brujin notation) during the reconstruction of the proof.
A lengthy proof shows that this preserves provability (we do not go into the details):
\begin{isabelle}%
\textsf{wf-theory}\ {\isasymTheta}\ {\isasymand}\ \textsf{finite}\ {\isasymGamma}\ {\isasymand}\ {\isacharparenleft}{\isasymforall}A{\isasymin}{\isasymGamma}{\isachardot}\ \textsf{wt-term}\ {\isacharparenleft}\textsf{sig}\ {\isasymTheta}{\isacharparenright}\ A\ {\isasymand}\ {\isasymturnstile}\isactrlsub {\isasymtau}\ A\ {\isacharcolon}\ \textsf{prop}{\isacharparenright}\ {\isasymand}\ {\isasymTheta}{\isacharcomma}{\isasymGamma}\ {\isasymturnstile}\ t\ {\isasymand}\ \textsf{beta-eta-norm}\ t\ {\isacharequal}\ \textsf{Some}\ u\ {\isasymlongrightarrow}\ {\isasymTheta}{\isacharcomma}{\isasymGamma}\ {\isasymturnstile}\ u%
\end{isabelle}
Isabelle's code generator needs some help handling the maps used in the (order-sorted) signatures.
We provide a refinement of maps to association lists.
Another problematic point is the definition of the type instance relation \isa{{\isacharparenleft}{\isasymlesssim}{\isacharparenright}}, which contains an (unbounded) existential
quantifier. To make this executable, we provide an implementation which tries to compute a suitable type 
substitution. In another step, we refine the type substitution to an association list as well.

In the end we obtain a proof checker
\begin{isabelle}%
\textsf{check-proof}\ {\isasymTheta}\ P\ p\ {\isacharequal}\ {\isacharparenleft}\textsf{wf-theory}\ {\isasymTheta}\ {\isasymand}\ \textsf{replay}\ {\isasymTheta}\ P\ {\isacharequal}\ \textsf{Some}\ p{\isacharparenright}%
\end{isabelle}
that checks theory \isa{{\isasymTheta}} and checks if proof \isa{P} proves the given
proposition \isa{p}. The latter check is important because the Isabelle theorems that we check
contain both a proof and a proposition that the theorem claims to prove. Function \isa{\textsf{check-proof}}
checks this claim.
As one of our main results, we can prove the correctness of our checker:
\begin{theorem}\label{check_with_hyps_sound}
\isa{\textsf{check-proof}\ {\isasymTheta}\ P\ p\ {\isasymlongrightarrow}\ {\isasymTheta}{\isacharcomma}\textsf{set}\ {\isacharparenleft}\textsf{hyps}\ P{\isacharparenright}\ {\isasymtturnstile}\ p}
\end{theorem}
The proof itself is conceptually simple and proceeds by induction over the structure of proof terms.
For each proof constructor we need to show that the corresponding inference rule leads to the same conclusion
as its functional version used by \isa{\textsf{replay}}.
Most of the proof effort goes into a large library of results about terms, types, signatures, substitutions,
wellformedness etc.\ required for the proof, most importantly the fact that \isa{{\isasymbeta}{\isasymeta}} normalization preserve provability.

\section{Size and Structure of the Formalization}

All material presented so far has been formalized in Isabelle/HOL. The definition of the inference system
(incl.\ types, terms etc.) resides in a separate theory \isa{Core} that depends only on the basic library of
Isabelle/HOL. It takes about 300 LOC and is fairly high level and readable -- we presented most of it.
This is at least an order or magnitude smaller than Isabelle's
inference kernel (which is not clearly delineated) -- of course the latter is optimized for performance.
Its abstract type of theorems alone
takes about 2,500 LOC, not counting any infrastructure of terms, types, unification etc.

The whole formalization consists of 10,000 LOC. The main components are:
\begin{itemize}
\item Almost half the formalization (4,700 LOC) is devoted to providing a library of operations on types and terms 
and their properties. This includes, among others, executable functions for type checking, different 
types of substitutions, abstractions, the wellformedness checks and \isa{{\isasymbeta}} and \isa{{\isasymeta}} reductions. 
\item Proving derived rules of our inference system takes up 3,000 LOC. A large part of this is deriving rules
for equality and the \isa{{\isasymbeta}} and \isa{{\isasymeta}} reductions. Weakening rules are also derived.
\item Making the wellformedness checks for (order-sorted) signatures 
and theories as well as the type instance checks executable takes 1,800 LOC.
\item Definition and correctness proof for the checker builds on the above material and take only
about 500 additional LOC.
\end{itemize}

\section{Integration with Isabelle}
\label{Integration}

As explained above, Isabelle generates SML code for the proof checker. This code has its own definitions
of types, terms etc.\ and needs to be interfaced with the corresponding data structures in Isabelle.
This step requires 150 lines of handwritten SML code (\emph{glue code}) that translates Isabelle's data structures
into the corresponding data structures in the generated proof checker such that we can feed them into
\isa{\textsf{check-proof}}. We cannot verify this code and therefore aim to keep it as small and simple as possible.
This is the reason for the previously mentioned \emph{intentional implementation bias} we introduced in our formalization.
We describe now how the various data types are translated. We call a translation trivial
if it merely replaces one constructor by another, possibly forgetting some information.

The translation of types and terms is trivial as their structure is almost identical in the two
settings. For Isabelle code experts it should be mentioned that the two \texttt{term}
constructors \texttt{Free} and \texttt{Var} in Isabelle (which both represent free variables
but \texttt{Var} can be instantiated by unification) are combined in type \isa{var}
of the formalization which we left unspecified but which in fact
looks like this: \isa{\isacommand{datatype}\ var\ {\isacharequal}\ Free\ name\ {\isacharbar}\ Var\ indexname}. This is purely to trivialize the glue code,
in our formalization \isa{var} is totally opaque.

Proof term translation is trivial except for two special cases.
Previously proved lemmas become axioms in the translation (see also below)
and so-called ``oracles'' (typically the result of unfinished proofs, i.e. ``sorry'' on the user level)
are rejected (but none of the theories we checked contain oracles).
Also remember that the translation of proofs is not safety critical because all that matters
is that in the end we obtain a correct proof of the claimed proposition.

We also provide functions to translate relevant content from the background theory: 
axioms and (order-sorted) signatures. %
This mostly amounts to extracting association lists from efficient internal data structures.
Translating the axioms also involves translating some alternative internal representation
of type class constraints into their standard form presented in Sect.~\ref{sec:TypeClassReasoning}.

The checker is integrated into Isabelle by calling it every time a new named theorem has been proved.
The set of theorems proved so far is added to the axiomatic basis for this check.
Cyclic dependencies between lemmas are ruled out by this ordering because every theorem is checked
before being added to the axiomatic basis. However, an explicit cyclicity check is not
part of the formalization (yet), which speaks only about checking single proofs.

\section{Running the Proof Checker}

We run this modified Isabelle with our proof checker on multiple theories in various object logics contained
in the Isabelle distribution.
A rough overview of the scope of the covered material for some logics and the required running times can be
found in the following table.
The running times are the total times for running Isabelle, not just the proof checking, but the latter
takes 90\% of the time. All tests were performed on a Intel Core i7-9750H CPU running at 2.60GHz and 32GB of RAM.

\begin{center}
\setlength\tabcolsep{1em}
\begin{tabular}{|l|r|l|}
\hline
Logic & LOC & Time \\\hline
FOL & ~4,500 & 45 secs\\
ZF & 55,000 & 25 mins\\
HOL & 10,000 & 26 mins\\
\hline
\end{tabular}
\end{center}

We can check the material in several smaller object logics in their entirety.
One of the larger such logics is first-order logic (FOL).
These logics do not develop any applications but FOL comes with proof automation and
theories testing that automation, in particular Pelletier's collection
of problems that were considered challenges in their day \cite{Pelletier}.
Because the proofs are found automatically, the resulting proof terms will typically be
quite complex and good test material for a proof checker.

The logic ZF (Zermelo-Fraenkel set theory) builds on FOL but contains real applications
and is an order of magnitude larger than FOL.
We are able to check all material formalized in ZF in the Isabelle distribution.

Isabelle's most frequently used and largest object logic is HOL.
We managed to check about 12\% of the \texttt{Main} library.
This includes the basic logic and the
libraries of sets, functions, orderings, lattices and groups.
The formalizations are non-trivial and make heavy use of Isabelle's type classes.

Why can we check about five times as many lines of code in ZF compared to HOL?
Profiling revealed that the proof checker spends a lot of time in functions that access the signature,
especially the wellformedness checks.
The primary reasons: inefficient data structures (e.g.\ association lists) and thus the running time depends heavily
on size of signature and increases with every new constant, type and class. To make matters worse,
there is no sharing of any kind in terms/types and their wellformedness checks.
Because ZF is free of polymorphism and type classes, these wellformedness checks are much simpler.

\section{Trust Assumptions}
\label{sec:Trust}

We need to trust the following components outside of the formalization:
\begin{itemize}
\item The verification (and code generation) of our proof
checker in Isabelle/HOL. This is inevitable, one has to trust some
theorem prover to start with. We could improve the trustworthiness of this step
by porting our proofs to the verified HOL prover by Kumar \emph{et el.} \cite{jar/KumarAMO16}
but its code generator produces CakeML \cite{CakeML}, not SML.
\item The unverified glue code in the integration of our proof checker into Isabelle
(Sect.~\ref{Integration}).
\end{itemize}

Because users currently cannot examine Isabelle's internal data structures that we start from,
they have to trust Isabelle's front end that parses and transforms some textual input file into
internal data structures. One could add a (possibly verified) presentation layer
that outputs those internal representations into a readable format that can be inspected, while avoiding
the traps Adams \cite{itp/Adams16} is concerned with.

\section{Future Work}

Our primary focus will be on scaling up the proof checker to not just deal with all of HOL but with
real applications (including itself!). There is a host of avenues for exploration. Just to name a few
promising directions:
more efficient data structures than association lists
(e.g.\ via existing frameworks \cite{Collections,Containers});
caching of wellformedness checks for types and terms;
exploiting sharing within terms and types (tricky because our intentionally simple glue code
creates copies); working with the compressed proof terms \cite{BerghoferN-TYPES00}
that Isabelle creates by default instead of uncompressing them as we do now.

We will also upgrade the formalization of our checker from individual theorems sets of theorems,
explicitly checking cyclic dependencies
(which are currently prevented by the glue code, see Sect.~\ref{Integration}).

A presentation layer as discussed in Sect.~\ref{sec:Trust} would not just allow the inspection
of the internal representation of the theories but could also be extended to the proofs themselves,
thus permitting checkers to be interfaced with Isabelle on a textual level instead of internal
data structures.

It would also be nice to have a model-theoretic semantics for \M. We believe that the work by
Kun\v{c}ar and Popescu \cite{itp/Kuncar015,esop/Kuncar017,pacmpl/Kuncar018,jar/KuncarP19a}
could be adapted from HOL to \M. This would in particular yield semantically justified
cyclicity checks for constant and type definitions which we currently treat as axioms
because a purely syntactic justification is unclear.

\subsection*{Acknowledgements}

We thank Kevin Kappelmann, Magnus Myreen, Larry Paulson, Andrei Popescu, Makarius Wenzel
and the anonymous reviewers for their comments.

\appendix
\section{Appendix}

\begin{isabelle}%
\textsf{subst-bv}\ u\ t\ {\isacharequal}\ \textsf{subst-bv2}\ t\ {\isadigit{0}}\ u%
\end{isabelle}

\begin{isabelle}%
\textsf{subst-bv2}\ {\isacharparenleft}\textsf{Bv}\ i{\isacharparenright}\ n\ u\ {\isacharequal}\ {\isacharparenleft}\textsf{if}\ i\ {\isacharless}\ n\ \textsf{then}\ \textsf{Bv}\ i\ \textsf{else}\ \textsf{if}\ i\ {\isacharequal}\ n\ \textsf{then}\ u\ \textsf{else}\ \textsf{Bv}\ {\isacharparenleft}i\ {\isacharminus}\ {\isadigit{1}}{\isacharparenright}{\isacharparenright}\isasep\isanewline%
\textsf{subst-bv2}\ {\isacharparenleft}\textsf{Abs}\ T\ t{\isacharparenright}\ n\ u\ {\isacharequal}\ \textsf{Abs}\ T\ {\isacharparenleft}\textsf{subst-bv2}\ t\ {\isacharparenleft}n\ {\isacharplus}\ {\isadigit{1}}{\isacharparenright}\ {\isacharparenleft}\textsf{lift}\ u\ {\isadigit{0}}{\isacharparenright}{\isacharparenright}\isasep\isanewline%
\textsf{subst-bv2}\ {\isacharparenleft}f\ {\isasymbullet}\ t{\isacharparenright}\ n\ u\ {\isacharequal}\ \textsf{subst-bv2}\ f\ n\ u\ {\isasymbullet}\ \textsf{subst-bv2}\ t\ n\ u\isasep\isanewline%
\textsf{subst-bv2}\ t\ \_\ \_\ {\isacharequal}\ t%
\end{isabelle}

\begin{isabelle}%
\textsf{lift}\ {\isacharparenleft}\textsf{Bv}\ i{\isacharparenright}\ n\ {\isacharequal}\ {\isacharparenleft}\textsf{if}\ n\ {\isasymle}\ i\ \textsf{then}\ \textsf{Bv}\ {\isacharparenleft}i\ {\isacharplus}\ {\isadigit{1}}{\isacharparenright}\ \textsf{else}\ \textsf{Bv}\ i{\isacharparenright}\isasep\isanewline%
\textsf{lift}\ {\isacharparenleft}\textsf{Abs}\ T\ t{\isacharparenright}\ n\ {\isacharequal}\ \textsf{Abs}\ T\ {\isacharparenleft}\textsf{lift}\ t\ {\isacharparenleft}n\ {\isacharplus}\ {\isadigit{1}}{\isacharparenright}{\isacharparenright}\isasep\isanewline%
\textsf{lift}\ {\isacharparenleft}f\ {\isasymbullet}\ t{\isacharparenright}\ n\ {\isacharequal}\ \textsf{lift}\ f\ n\ {\isasymbullet}\ \textsf{lift}\ t\ n\isasep\isanewline%
\textsf{lift}\ t\ \_\ {\isacharequal}\ t%
\end{isabelle}

\begin{isabelle}%
\textsf{bind-fv}\ T\ t\ {\isacharequal}\ \textsf{bind-fv2}\ T\ {\isadigit{0}}\ t%
\end{isabelle}

\begin{isabelle}%
\textsf{bind-fv2}\ var\ n\ {\isacharparenleft}\textsf{Fv}\ v\ T{\isacharparenright}\ {\isacharequal}\ {\isacharparenleft}\textsf{if}\ var\ {\isacharequal}\ {\isacharparenleft}v{\isacharcomma}\ T{\isacharparenright}\ \textsf{then}\ \textsf{Bv}\ n\ \textsf{else}\ \textsf{Fv}\ v\ T{\isacharparenright}\isasep\isanewline%
\textsf{bind-fv2}\ var\ n\ {\isacharparenleft}\textsf{Abs}\ T\ t{\isacharparenright}\ {\isacharequal}\ \textsf{Abs}\ T\ {\isacharparenleft}\textsf{bind-fv2}\ var\ {\isacharparenleft}n\ {\isacharplus}\ {\isadigit{1}}{\isacharparenright}\ t{\isacharparenright}\isasep\isanewline%
\textsf{bind-fv2}\ var\ n\ {\isacharparenleft}f\ {\isasymbullet}\ u{\isacharparenright}\ {\isacharequal}\ \textsf{bind-fv2}\ var\ n\ f\ {\isasymbullet}\ \textsf{bind-fv2}\ var\ n\ u\isasep\isanewline%
\textsf{bind-fv2}\ \_\ \_\ t\ {\isacharequal}\ t%
\end{isabelle}

\bibliographystyle{splncs04}
\bibliography{root}%
\end{isamarkuptext}\isamarkuptrue%
\isadelimtheory
\endisadelimtheory
\isatagtheory
\endisatagtheory
{\isafoldtheory}%
\isadelimtheory
\endisadelimtheory
\end{isabellebody}%

\end{document}